\documentclass[aps,showpacs]{revtex4-2}
\usepackage{graphicx}
\usepackage{color}
\usepackage{slashed}
\def\bc{\begin{center}}
\def\ec{\end{center}}

\def\beq{\begin{equation}}
\def\eeq{\end{equation}}

\def\br{{\bf r}}
\def\bp{{\bf q}}
\def\bk{{\bf k}}
\def\bp{{\bf p}}

\def\bx{{\bf x}}

\def\vK{{\vec K}}
\def\sgn{{\rm sgn}}

\begin{document}

\title{
Phonon mode splitting and phonon anomaly in multiband electron systems
}

\author{K. Ziegler
}
\affiliation{Institut f\"ur Physik, Universit\"at Augsburg, D-86135 Augsburg, Germany\\
}
\affiliation{
Physics Department, New York City College of Technology,\\
The City University of New York,
Brooklyn, NY 11201, USA}

\date{\today}

\begin{abstract}
We investigate the topological consequences of coupling chiral fermions to local, dispersionless phonons. This interaction induces a splitting of the phonon spectrum into three bands: a flat band and two bands with linear dispersion, all of which are degenerate at a nodal point located at zero wavevector.
The flat band exhibits vanishing Berry curvature, while the linearly dispersing bands carry nontrivial topological features. Their Berry curvature fields assume a hedgehog-like structure in momentum space, analogous to monopole configurations, and reflect the chirality of the underlying fermionic system.
Moreover, the effective phonon response reveals a phonon parity anomaly, observable as a discontinuity in the phonon current. This anomaly originates from the singularities of the fermion Green’s function and signals the transfer of topological information from fermions to phonons.
Our results demonstrate that phonon currents provide a direct probe of electronic chirality and topological structures. 


\end{abstract}

\maketitle


\section{Introduction}

Phonons, the quantized modes of lattice vibrations, are the primary carriers of thermal energy in insulating and
semiconducting solids. Understanding their transport behavior is essential for describing and engineering thermal 
conductivity in materials. Traditional approaches to phonon transport rely on the Boltzmann transport 
equation~\cite{callaway59,hardy63,mahan90}, where either harmonic or anharmonic lattice models were
studied.
In thermal transport theory, using the Boltzmann transport equation,
the phonon current density $\bf{j}_{\rm phonon}$ can be written in the form~\cite{hardy63}
\beq
\label{ph_current00}
\bf{j}_{\rm{phonon}} = \frac{1}{V} \sum_{\bk,s} \hbar \omega_{\bk,s}\, \bf{v}_{\bk,s}\, 
\delta f_{\bk,s},
\eeq
where $\hbar \omega_{\bk,s}$ is the energy of a phonon mode $s$ with wave vector $\bk$
and $\bf{v}_{\bk,s}$ is the group velocity of that phonon. $\delta f_{\bk,s}$ represents 
the deviation of the phonon distribution from equilibrium, either caused by a temperature gradient or by other 
perturbations. Finally, $V$ is the volume of the material.
This expression quantifies how the non-equilibrium distribution of phonons leads to an energy current 
across the material. Note that although phonons are not conserved (they can be created or annihilated 
in interactions), the energy they carry is what is crucial for processes like thermal conductivity. 
We note that the current is parallel to the group velocity $\bf{v}_{\bk,s}$ in this approach;
i.e., there is no Hall effect. This changes when we apply either a magnetic field or include spins,
which are coupled to the lattice. 
%
This can cause the anomalous Hall effect for phonons. It was initially observed for
electronic systems but later also in an experiment with phonons~\cite{strohm05}. 
Kagan and Maximov studied the effect of phonons coupled to a spin system and found a transverse transport
coefficient, which arises due to spin-phonon interaction~\cite{kagan08}. 
Another example for the anomalous Hall effect was observed for an ionic crystal lattice 
in the presence of a static magnetic field~\cite{zhang10}, based on a model with harmonic coupling~\cite{holz71}.  
It was demonstrated that phonon modes possess a Berry curvature, indicating a geometric origin for a
transverse thermal response.
Other examples of topological phonon effects in different lattice models and materials have been
studied recently, ranging from magnon-phonon systems in honeycomb ferromagnets~\cite{thingstad19},
to graphene and similar lattice materials as well as waveguides~\cite{singh18,liu20,li20a,li20,ding23,xi25}.
A remaining issue is that the hallmark of the electronic quantum Hall effect, namely the conductivity plateaux,  
has not been observed in the phononic transport. This is the motivation of the present work, in which we will
analyze (i) whether a transverse phonon current is induced by phonons that are coupled to another system
and (ii) what the corresponding phonon conductivity is. In this context we are interested in general results that
can be applied to specific systems subsequently.
Very important are systems, where the phonons couple to electrons. 
Electron-phonon interaction is the origin of many fundamental phenomena in condensed matter physics.
For instance, it is central to the foundation of the BCS theory for 
superconductivity~\cite{bardeen57} and has many other applications, such as the Jahn-Teller 
effect~\cite{jahn37}. The effect of the electron-phonon interaction on the electronic transport properties
in two-dimensional systems was studied for (massive) Dirac fermions~\cite{li13} and for a system with
broken inversion symmetry~\cite{yar24,zhang26}.
Another effect is the coupling of phonons to chiral fermions, which can lead to the splitting of phonon 
modes~\cite{ziegler11,ziegler11a}, to the renormalization of the effective phonon Green's function~\cite{basko08} 
as well as to the electronic quantum Hall effect~\cite{sinner16,sinner19,sinner20}. The emergence of chiral
phonons was also studied for electrons on the Kagome lattice~\cite{chen19,chen25}.
Inspired by these examples, we will investigate in the following the phonon properties that emerge 
in systems with mixed even and odd parity. The primary goal of this paper is to explore mechanisms by which 
Dirac-like or topologically nontrivial phonon modes can arise. Specifically, we examine how an initially isotropic
and disordered phonon field may split into multiple, energetically distinct and dispersive phonon modes through 
interactions with fermions or other degrees of freedom. A central question we seek to answer is what topological 
or spectral characteristics do phonons inherit from the fermions they couple to?

The paper is organized as follows: In Sect. \ref{sect:model} the model for electron-phonon interaction is defined.
Based on this model the phonon modes are treated within a functional-integral representation in 
Sect. \ref{sect:functional_int}, which provides an effective action resembling a Chern-Simons term for 2+1
dimensions. The phonon current is calculated in Sect. \ref{sect:currents} and the phonon eigenbasis is
studied in Sect. \ref{sect:eigenbasis}, including a discussion of the corresponding Berry curvatures. 
These results are discussed in Sect. \ref{sect:discussion}, deriving a phonon anomaly for chiral fermions
(Sect. \ref{sect:weyl}) and the absence of the phonon anomaly for non-chiral fermions (Sect. \ref{sect:non-chiral}).
Sect. \ref{sect:conclusion} concludes the results of this work and gives an outlook to open questions.

\section{Model}
\label{sect:model}

The phonon Hamiltonian with three directions $\lambda=1,2,3$ of the atomic displacements reads
\beq
\label{hamiltonian_p}
{\hat H}_{\rm ph}=\sum_\br\sum_{\lambda=1}^3b^\dagger_{\br,\lambda}\omega_{\br\br'} b^{}_{\br',\lambda}
+h.c.
\eeq
with bosonic creation and annihilation operators $b^\dagger_{\br,\lambda}$ and $b_{\br,\lambda}$ for a site $\br$
on a two-dimensional lattice. Three independent directions of atomic displacements are considered, motivated by
two in-plane and one out-off-plane modes in a honeycomb lattice~\cite{basko08,ziegler11,ziegler11a}.
The case of two in-plane modes was studied previously~\cite{sinner16,sinner19,sinner19}, while the extension
to the full $SU(4)$ group of formal displacements was discussed in a more recent work~\cite{sinner20}.
For simplicity, we assume uncorrelated local phonon fluctuations by choosing 
$\omega_{\br\br'} =\omega_0\delta_{\br\br'}$. The hopping Hamiltonian of chiral electrons, created and
annihilated by $c^\dagger_{\br\mu}$ and $c_{\br\mu}$, with $N$ bands reads
\beq
\label{hamiltonian_e}
{\hat H}_{\rm el}=\sum_{\br,\br\rq{}}\sum_{\lambda=1}^3\sum_{\mu,\mu\rq{}=1}^N
c^\dagger_{\br\mu}h_{\br-\br',\lambda}\gamma_{\lambda;\mu\mu'} c_{\br'\mu'} +h.c.
\eeq
with generalized Pauli matrices $\{\gamma_\lambda\}$. 
Here we note that a model with $N$ bands would be defined by three $N\times N$ matrices $\{\gamma_\lambda\}$,
since we restrict the following discussion to two spatial dimensions 
and the time axis. The Hamiltonian may have only two non-zero coefficients, describing the tunneling of
electrons on a two-dimensional lattice.
Although we do not specify these matrices explicitly, we can always replace them by Pauli matrices for two bands.
This enables us to interpret this model as a tight-binding Hamiltonian on a honeycomb lattice.
The hopping terms determine the behavior of the Hamiltonian under parity transformation.
We assume that at least some are odd under parity transformation: $h_{\br-\br',\lambda}=-h_{\br'-\br,\lambda}$.
Finally, the Hamiltonian of electron-phonon Holstein coupling~\cite{holstein59} is
\beq
\label{hamiltonian_ep}
{\hat H}_{\rm el-ph}=g\sum_{\br}\sum_{\lambda=1}^3\sum_{\mu,\mu'=1}^N\hat{Q}_{\br,\lambda}
c^\dagger_{\br\mu}\gamma_{\lambda;\mu\mu'} c_{\br\mu'}
+h.c.
\eeq
with the phonon operator $\hat{Q}_{\br,\lambda}=b^\dagger_{\br,\lambda}+b_{\br,\lambda}$. 
$g$ is the electron-phonon coupling strength, where $g=0$ decouples phonons from electrons.
A third matrix $\gamma_3$ appears in the interaction with the displacement $\hat{Q}_{\br,3}$,
representing an out-of-plane displacement in the case of graphene~\cite{ziegler11}.
In the following we consider the electron-phonon system that is described by the Hamiltonian 
${\hat H}={\hat H}_{\rm ph}+{\hat H}_{\rm el}+{\hat H}_{\rm el-ph}$.
For the subsequent calculations it is convenient to employ the functional-integral representation
of the partition function, which will be discussed in the next section. For instance, the
time-dependent correlation of phonons can be defined as a functional integral.

\subsection{Functional-integral representation}
\label{sect:functional_int}

We start from the functional integral representation~\cite{negele} of the phonon correlator for $\bx=(\br,t)$
\beq
\label{f-int}
\frac{1}{Z}\int Q_\bx Q_{\bx'}\exp(-S){\cal D}[\psi,Q]
\ \ {\rm with}\ \ 
Z =\int\exp(-S){\cal D}[\psi,Q]
\eeq
with the covariant action~\cite{sinner16,sinner19,sinner20} 
\beq
S=
\int_\bx
\left(\omega_0Q_\bx^2+{\bar\psi}_{\bx}i\gamma_3\partial_t\psi_{\bx}+ H_{\rm el-ph}[\psi_{\bx},Q_{\bx}]\right)
,
\eeq
where ${\bar\psi}_\bx$ is the complex conjugate of the Grassmann field 
$\psi_\bx$. $H_{\rm el-ph}[\psi_{\bx},Q_{\bx}]$ is a functional of the fields, obtained from the 
Hamiltonian ${\hat H}_{\rm el-ph}$, where the fermion operators $c_{\br\mu}$,  $c^\dagger_{\br\mu}$ are 
replaced by the Grassmann vector fields $\psi_\bx$, $\psi^*_\bx$, respectively and the phonon operator 
$\hat{Q}_{\br,\lambda}$ is 
replaced by a real vector field $Q_\bx\equiv(Q_{\bx,1},Q_{\bx,2},Q_{\bx,3})$. 
The functional integral in Eq. (\ref{f-int}) can be used to calculate thermal fluctuations by an
analytic continuation of the time $t$ integration to the interval $[0,i\beta]$, where $\beta$ is the inverse
temperature $\beta=1/k_BT$~\cite{negele}.

After integrating over the Grassmann field and rescaling $Q_\bx\to gQ_\bx$
we get an effective action for the phonon distribution as
\beq
S_Q=\frac{\omega_0}{g^2}\sum_{\lambda=1}^3\int_\bx Q_{\bx,\lambda}^2  
+\int_\bx Tr_{N}
\log G^{-1}
\eeq
with the inverse fermion Green's function 
$G^{-1}=(h_1+Q_1)\gamma_1+(h_2+Q_2)\gamma_2+(i\partial_t+Q_3)\gamma_3+im\gamma_0$.
The scalar parameter $im$ serves as an imaginary energy term that regularizes the Green's function.
In principle, it can be sent to zero in the end of the calculation. We will see later that it plays a crucial
role in terms of a phonon anomaly.
The structure of $G^{-1}$ resembles the Dirac operator with gauge field $Q_\mu$
for Dirac matrices $\{\gamma_\nu\}$ and unit matrix $\gamma_0$~\cite{adler69,redlich84,witten16} . 
In particular, for 2+1 dimensional chiral (Weyl) fermions the $\{\gamma_\nu\}$ are Pauli matrices.
The first integral describes uncorrelated random phonons, the second integral the electron-phonon
coupling.

Next we search for stationary solutions of the saddle-point equation $\delta_{Q_\nu} S_Q=0$.
An extended discussion of possible solutions was given in Ref. \cite{sinner20}. 
We will not repeat the calculation for specific solutions here but discuss the general properties of such solutions. 
Assuming that $\gamma_\nu\gamma_{\nu'}+\gamma_\nu\gamma_{\nu'}=0$ for $\nu'\ne\nu$,
there is always a solution ${\bar Q}=0$, whose validity is justified by the stability of the fluctuations 
$q_\bx=Q_\bx-{\bar Q}$.
If this solution is unstable, though, we must search for ${\bar Q}\ne 0$.
For the following we consider small values of $g$ to stay in the stable regime for fluctuations around ${\bar Q}_\nu=0$.
Once we have found a saddle point ${\bar Q}$, 
we expand $S_Q$ around it up to second order in the fluctuating field $q_\bx$:
\beq
\label{gaussian0}
S_Q=S_{{\bar Q}}+\frac{\omega_0}{g^2}\sum_\mu\int_\bx q_{\bx\mu}^2 
-\frac{1}{2}\int_\bx\int_{\bx'} Tr_{N}(q_\bx G_{\bx-\bx'}q_{\bx'}G_{\bx'-\bx})   
+...
\ ,
\eeq
where the first order term in $q_\bx$ vanishes due to the saddle point equation. 
Then we expand $q_\bx$
in terms of $\{\gamma_\mu\}_{\mu=1,2,3}$ and the unit matrix $\gamma_0$ as
\beq
q_\bx=\sum_{\nu=1}^3 q_{\bx\nu}\gamma_\nu
\ ,\ \ 
G_{{\bar\bx}}=\sum_{\nu=0}^3 G_{{\bar\bx}\nu}\gamma_\nu
,
\eeq
where only the Green's function includes $\nu=0$.
After inserting this into $S_Q$, the leading term for the long-range behavior
is the linear differential term (cf. App. \ref{app:CS}):
\beq
\label{cs_action}
S_{1}=\frac{\omega_0}{g^2}\sum_\mu\int_\bx q_{\bx\mu}^2 
+\sum_{\mu,\mu'}D_{\mu\mu'}\int_\bx q_{\bx\mu}q_{\bx\mu'}
+\sum_{\lambda,\mu,\mu'=1}^3\Gamma_{\lambda;\mu\mu'}
\int_\bx q_{\bx\mu}\frac{\partial q_{\bx\mu'}}{\partial x_\lambda} 
\eeq
with the tensors
\beq
\label{Gamma_tensor}
D_{\mu\mu'}=\frac{\omega_0}{g^2}\delta_{\mu\mu'}-\frac{1}{2}\sum_{\nu,\nu'=0}^3
\int_\bk{\tilde G}_{\bk\nu}{\tilde G}_{\bk\nu'}Tr_N(\gamma_{\mu}\gamma_{\nu}\gamma_{\mu'}\gamma_{\nu'})
\ ,\ \ 
\Gamma_{\lambda;\mu\mu'}=\sum_{\nu,\nu'=0}^3C_{\lambda;\nu,\nu'} \tau_{\mu\nu\mu'\nu'}
,
\eeq
where the second tensor consists of (cf. Eq. (\ref{antisymmetric1}))
\beq
\label{tensor0}
C_{\lambda;\nu,\nu'}
=\frac{1}{4}\int_\bx x_\lambda G_{{\bx}\nu} G_{-{\bx}\nu'} 
=\frac{i}{4}\int_\bk {\tilde G}_{\bk\nu}\frac{{\partial {\tilde G}_{\bk \nu'}}}{\partial k_\lambda} 
\ ,\ \ 
\tau_{\mu\nu\mu'\nu'}=
Tr_{N}(\gamma_{\mu}\gamma_{\nu}\gamma_{\mu'}\gamma_{\nu'}
-\gamma_{\mu'}\gamma_{\nu}\gamma_{\mu}\gamma_{\nu'})
.
\eeq
We note that $\tau_{\mu\nu\mu'\nu'}=-\tau_{\mu'\nu\mu\nu'}$, which implies that $\Gamma_{\lambda;\mu\mu'}$
is antisymmetric in $\mu$ and $\mu'$. 
This property removes the boundary terms of the partial integration in Eq. (\ref{cs_action}) due to
\beq
\sum_{\mu,\mu'=1}^3  \tau_{\mu\nu\mu'\nu'}q_{\bx\mu'}q_{\bx\mu}\Big|_{\rm boundary}=0
.
\eeq
Moreover, in the spatial integral of $C_{\lambda;\nu\nu'}$  in Eq. (\ref{tensor0}) the integrand
must have an even parity for a non-vanishing integral, which requires that the product 
$G_{{\bx}\nu} G_{-{\bx}\nu'}$ must be odd under parity transformation. Thus, the Green's function
$G$ must consist of a sum of terms with odd and even parity.
Finally, we
assume that $D_{\mu\mu'}={\bar D}\delta_{\mu\mu'}$, which is justified for the examples we will
discuss in Sect.\ref{sect:discussion}.

The kernel of the quadratic form in Eq. (\ref{cs_action}) can be understood as an 
inverse phonon Green's function
\beq
\label{phonon_ham}
{\cal K}^{-1}_{\mu\mu'}=\Omega_0\delta_{\mu\mu'}
+\sum_{\lambda=1}^3 \Gamma_{\lambda;\mu\mu'}\frac{\partial}{\partial x_\lambda}
\ ,\ \
\Omega_0=\frac{\omega_0}{g^2}-{\bar D}
,
\eeq
where the second term resembles an inverse chiral (Weyl) Green's function, whose Pauli matrices are replaced by the 
$3\times3$ matrices $\{\Gamma_{\lambda}\}$. 
Due to the antisymmetry there are only three independent tensor elements for each value of $\lambda$.
This suggests the linear mapping in coordinate space
\beq
\label{coordinate_tr}
\cases{
\partial_{x'_1}=-\sum_\lambda\Gamma_{\lambda;23}\partial_{x_\lambda} \cr
\partial_{x'_2}=\sum_\lambda\Gamma_{\lambda;13}\partial_{x_\lambda} \cr
\partial_{x'_3}=-\sum_\lambda\Gamma_{\lambda;12}\partial_{x_\lambda} \cr
}
\eeq
such that we get for the inverse chiral Green's function in Eq. (\ref{phonon_ham}) 
\beq
\label{phonon_ham2}
\sum_{\lambda=1}^3 \Gamma_{\lambda;\mu\mu'}\frac{\partial}{\partial x_\lambda}
=\sum_{\lambda=1}^3\epsilon_{\mu\lambda\mu'}\frac{\partial}{\partial x'_\lambda}
.
\eeq
This relation indicates that the phonon Green's function can always be mapped onto a conventional
2+1 dimensional Chern-Simons term~\cite{witten16}, where the map itself carries the information about the fermion
system, to which the phonons are coupled. Thus, we have separated the properties of the fermion part
from the phonon part through the map in Eq. (\ref{coordinate_tr}), comprising the parameters 
$\Gamma_{\lambda;12}$, $\Gamma_{\lambda;13}$ and $\Gamma_{\lambda;23}$.
Besides this map, the inverse phonon Green's function is simply a conventional Chern-Simons term.

\section{Phonon currents}
\label{sect:currents}

Nonequilibrium phonon modes create phonon currents. These currents are calculated, 
in analogy with the gauge theory~\cite{witten16}, 
from the variation of the action in Eq. (\ref{cs_action})
as the linear response to the gradient of the phonon field: 
\beq
\label{current01}
j_\mu=\frac{1}{2}\delta_{q_{\bx\mu}} S_1
=\Omega_0q_{\bx\mu}+\frac{1}{2}
\delta_{q_{\bx\mu}}\sum_{\lambda,\mu,\mu'=1}^3\Gamma_{\lambda;\mu\mu'}
\int_\bx q_{\bx\mu}\frac{\partial q_{\bx\mu'}}{\partial x_\lambda} 
=\Omega_0q_{\bx\mu}
+\sum_{\lambda,\mu'=1}^3\Gamma_{\lambda;\mu\mu'}\frac{\partial q_{\bx\mu'}}{\partial x_\lambda}
\eeq
and after the coordinate transformation
\beq
j_\mu=\Omega_0q_{\bx\mu}
+\sum_{\lambda,\mu'=1}^3\epsilon_{\mu\lambda\mu'}\frac{\partial  q_{\bx\mu'}}{\partial x'_\lambda}
.
\eeq
Thus, we have a longitudinal current proportional to $\Omega_0$ and a universal transverse current, which
is induced by the gradient of the phonon field. Both terms vanish when $ q_{\bx\mu}$ is a uniform equilibrium
phonon mode, since $ q_{\bx\mu}=0$ in this case. For $ q_{\bx\mu}\ne 0$, on the other hand, we can create
longitudinal as well as transverse currents simultaneously.

Alternatively, we define the phonon current operator through the inverse phonon Green's function as
\beq
\label{current0}
i[{\cal K}^{-1},x_\lambda]
=i\Gamma_{\lambda}
.
\eeq
The phonon current expectation values with respect to a phonon field ${\vec\psi}=(\psi_1,\psi_2,\psi_3)^T$ 
and ${\vec\Gamma}_\lambda=(\Gamma_{\lambda;23},-\Gamma_{\lambda;13},\Gamma_{\lambda;12})^T$ read
\beq
\label{current2}
({\vec\psi}\cdot{\bf j}_\lambda{\vec\psi})
={\vec\Gamma}_\lambda\cdot({\vec\psi}^*\times{\vec\psi})
.
\eeq
The corresponding fermion current operator reads in Fourier representation
\beq
{\bf J}_\lambda=[{\tilde H}_{\bk},\partial_{k_\lambda}]=\sum_{\mu=1}^3
\frac{\partial h_{\bk,\mu}}{\partial k_\lambda}\gamma_\mu
\ ,\ \
{\tilde H}_{\bk}
=\sum_{\mu=1}^3 {\tilde h}_{\bk,\mu}\gamma_{\mu}
.
\eeq
While the fermion current is given by the $N\times N$ generalized Pauli matrices $\{\gamma_\mu\}$, 
the corresponding phonon current is related to the antisymmetric $3\times3$ matrices 
$\{\Gamma_\lambda\}$ of Eq. (\ref{Gamma_tensor}). 


\subsection{Phonon Hamiltonian eigenbasis}
\label{sect:eigenbasis}

\begin{figure*}[t]
\begin{center}
\includegraphics[height=7cm,width=5cm]{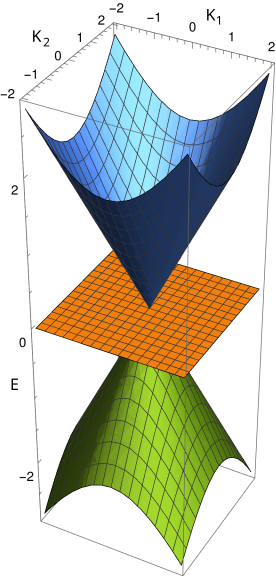}
\caption{
Three phonon bands plotted for $K_3=0$ with a node at $K_1=K_2=0$.
All three bands are shifted by the constant $-\Omega_0$.
}
\label{fig:a}
\end{center}
\end{figure*}

\begin{figure*}[t]
\begin{center}
\includegraphics[width=7.7cm]{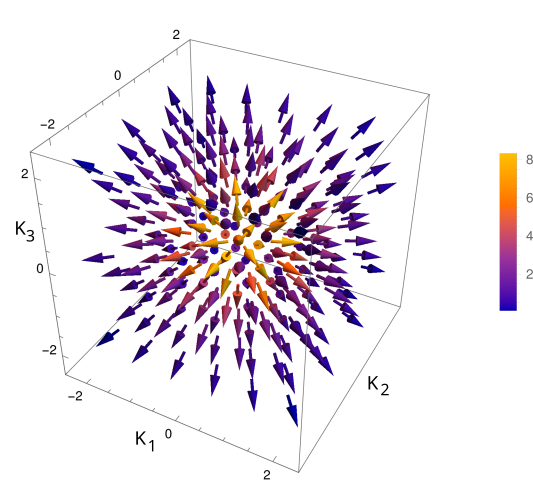}
\includegraphics[width=7.7cm]{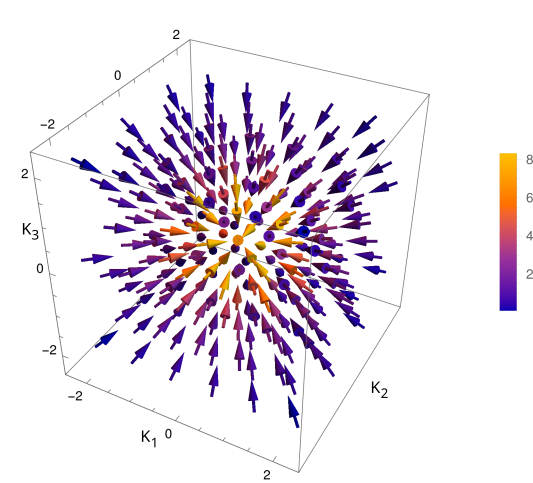}
\caption{
Hedgehog-like Berry curvature of the upper (left) and the lower band (right).
}
\label{fig:b}
\end{center}
\end{figure*}

So far, all vectors and tensors are written in terms of the natural basis, which was introduced in the
Hamiltonians in Eqs. (\ref{hamiltonian_p}) -- (\ref{hamiltonian_ep}). Next we study the eigenbasis
of the inverse phonon Green's function of Eq. (\ref{phonon_ham}) and its properties.
It is convenient to work in the transformed coordinate space $\bx'\equiv (x'_j)$, where we can diagonalize
the expression in Eq. (\ref{phonon_ham2}) after the Fourier transformation $x_j'\to K_j$
with the vector $\vK=-\sum_{\lambda}k_\lambda{\vec\Gamma}_\lambda$.
Then the eigenvalues of
\beq
\pmatrix{
\Omega_0 & -iK_3 & iK_2 \cr
iK_3 & \Omega_0 & -iK_1 \cr
-iK_2 & iK_1 & \Omega_0 \cr
}
\eeq
are $E_0=\Omega_0$ and $E_\pm=\Omega_0\pm |{\vec K}|$
with the corresponding normalized eigenvectors
\beq
\label{eigenvectors}
{\vec\psi}_0=-\frac{\vK}{|\vK|}
\ ,\ \ \
{\vec\psi}_\pm=\frac{1}{\sqrt{2}\sqrt{K_2^2+K_3^2}|{\vec K}|}\pmatrix{
K_2^2+K_3^2 \cr
-K_1K_2\pm iK_3|{\vec K}| \cr
-K_1K_3\mp iK_2|{\vec K}|\cr 
}
.
\eeq
Thus, we have a flat band and two bands with a linear dispersion. These three bands are degenerate at
the node $\vK=0$ (cf. Fig.\ref{fig:a}). 
There is no gap as long as the diagonal elements of the inverse phonon Green's function
are identical, even if the fermion Hamiltonian is gapped.
For these eigenvectors the Berry connections read 
\beq
-i({\vec\psi}_0\cdot\nabla_\vK{\vec\psi}_0)=0
\ ,\ \ 
-i({\vec\psi}_\pm\cdot\nabla_\vK{\vec\psi}_\pm)=\mp\frac{K_1}{|\vK|(K_2^2+K_3^2)}
\pmatrix{
0 \cr
K_3 \cr
-K_2 \cr
}
.
\eeq
While the Berry connection for ${\vec\psi}_0$ vanishes, the  Berry connections of ${\vec\psi}_\pm$
represent two opposite vortices in the $K_2$--$K_3$ plane.
The corresponding Berry curvatures of $\psi_\pm$ are
\beq
\label{Berry_curvature1}
{\bf B}=-\nabla_\vK\times i({\vec\psi}_\pm\cdot\nabla_\vK{\vec\psi}_\pm)=\mp\frac{1}{|\vK|^3}{\vK}
=\pm\nabla_\vK\frac{1}{|\vK|}
,
\eeq
which represent vector fields in the form of hedgehogs for the upper and the lower band with opposite 
orientation (Fig. \ref{fig:b}). The divergence $\nabla\cdot{\bf B}=\pm4\pi\delta^{(3)}(\vK)$
reflects that these vector fields represent two Dirac Delta sources with charges $\pm4\pi$
at $\vK=0$ (Dirac monopoles). 
Hedgehog solutions have also been discussed in the context of 2D quantum spin systems~\cite{senthil04,fradkin13}.
In contrast, the Berry curvature for two-dimensional chiral fermions is $\pi\delta^{(2)}(\bk)$. 
This reflects a mapping from the fermion Berry curvature to the phonon Berry curvature.

Returning to the current operator of Eq. (\ref{current2}), the latter is calculated in the eigenbasis of ${\cal K}$.
Since ${\vec\psi}_0$ are real, its current ${\vec\Gamma}\cdot({\vec\psi}_0^*\times{\vec\psi}_0)$ vanishes.
For the other two eigenvectors we obtain
\beq
\label{current1}
({\vec\psi}_\pm\cdot\Gamma_\lambda{\vec\psi}_\pm)
=\pm i\frac{K_1\Gamma_{\lambda;23}-K_2\Gamma_{\lambda;13}+K_3\Gamma_{\lambda;12}}{|\vK|}
=\mp \frac{i}{|\vK|}\sum_{\lambda'}\beta_{\lambda\lambda'}k_{\lambda'}
\eeq
with $\beta_{\lambda\lambda'}=\Gamma_{\lambda;12}\Gamma_{\lambda';12}
+\Gamma_{\lambda;13}\Gamma_{\lambda';13}+\Gamma_{\lambda;23}\Gamma_{\lambda';23}$.
The quadratic form in terms of the current matrix elements is caused by the specific form of ${\vec\psi}_\pm$
in Eq. (\ref{eigenvectors}), in contrast to the general linear dependence in Eq. (\ref{current2}).

\section{Discussion of the results}
\label{sect:discussion}

The properties of the phonon modes, described by the inverse Green's function in Eq. (\ref{phonon_ham}),
are determined by the antisymmetric tensor 
$\Gamma_{\lambda;\mu\mu'}$, defined in Eqs. (\ref{Gamma_tensor}) and (\ref{tensor0}).
This raises the question about the very existence of this tensor. And if it exists, what are its properties?
As already mentioned in Sect. \ref{sect:functional_int}, its existence is intimately related to the fact that
the fermion Green's function is neither even nor odd under parity transformation. 
We write $G^{-1}=A+imS$, where $A$ and $S$ are Hermitian, and $A$ is odd and $S$ is even under parity 
transformation.
$m$ is a real parameter that controls the contribution of the even part and $imS$ serves as a regularization
of the Green's function. We can apply an analytic continuation of $im$ and at the end of the calculation 
it should be sent to 0. 
A nonzero $m$ causes a broadening of the Green's function, which typically occurs in the presence
of random scattering~\cite{mahan90}.
After writing $G=(A+imS)^{-1}=[(A-imS)(A+imS)]^{-1}(A-imS)$, we assume that
$[A,S]=0$. Then $G$ can be split as $G=a-ims$ with the odd part $a=(A^2+m^2S^2)^{-1}A$ 
and the even part $s=(A^2+m^2S^2)^{-1}S$.
The Green's function is expanded in terms of the matrices $\{\gamma_\nu\}$
 as $G_\bx=\sum_\nu G_{\bx\nu}\gamma_\nu$.
This gives us with Eq. (\ref{Gamma_tensor}) the Hermitian expression
\beq
\label{Gamma2}
\Gamma_{\lambda;\mu\mu'}=-\frac{im}{2}\sum_{\nu,\nu'}\int_\bx x_\lambda
(a_{\bx\nu}s_{\bx\nu'}+s_{\bx\nu}a_{\bx\nu'})\tau_{\mu\nu\mu'\nu'}
,
\eeq
which switches its sign for $m\to-m$, since the integrand depends only on $m^2$.

In the next step we must estimate the $m^2$ dependence of the integral. This will be studied in terms
of the specific example $a=(H^2+m^2\gamma_0)^{-1}H$ and $s=(H^2+m^2\gamma_0)^{-1}$
with the Fourier components ${\tilde H}_{\bk;\mu\mu'}
=\sum_{\nu=1}^3 {\tilde h}_{\bk,\nu}\gamma_{\nu;\mu\mu'}$ of the fermion Hamiltonian.
This implies for the integral in Eq. (\ref{tensor0})
\beq
C_{\lambda;\nu\nu'}=
\]
\[
\frac{m}{4}\int_\bk\left([(H_\bk^2+m^2\sigma_0)^{-1}H_\bk]_\nu \partial_{k_\lambda}[(H_\bk^2+m^2\sigma_0)^{-1}]_{\nu'}
+[(H_\bk^2+m^2\sigma_0)^{-1}]_\nu\partial_{k_\lambda}[(H_\bk^2+m^2\sigma_0)^{-1}H_\bk]_{\nu'} 
\right)
,
\eeq
which reads after a partial integration for the second integral
\beq
\label{tensor2}
=\frac{1}{4}m\left\{\int_\bk[(H_\bk^2+m^2\sigma_0)^{-1}H_\bk]_\nu\partial_{k_\lambda}
[(H_\bk^2+m^2\sigma_0)^{-1}]_{\nu'}-(\nu\leftrightarrow\nu')
\right\}
.
\eeq
Assuming that ${\tilde h}_{\bk,\nu}$ is linear in $\bk$, there is a single node at $\bk=0$.
Then we absorb the $1/m^2$ factor by rescaling $\bk\to\bk/|m|=:\bp$ to get
\beq
\label{C_tensor2}
C_{\lambda;\nu\nu'}
=\frac{1}{4}\sgn(m)|m|^{d-3}
\left\{\int_\bp [(H_\bp^2+\sigma_0)^{-1}H_\bp]_\nu\partial_{p_\lambda}
[(H_\bp^2+\sigma_0)^{-1}]_{\nu'}-(\nu\leftrightarrow\nu')
\right\}
.
\eeq
We are primarily interested in the jump for $d=3$, i.e., in the vicinity of $m\sim0$. 
Due to $C_{\lambda;\nu\nu}=0$, we only consider $\nu'\ne\nu$. 
Since we have assumed that ${\tilde h}_{\bp,\nu}$ is linear in $\bp$, the derivative 
$\partial_{p_\lambda}H_\bp^2$ gives an expression that is linear in $\bp$.
In the product of Eq. (\ref{tensor2}) the factor $H_\bk$ is also linear in $\bk$, such that the integrand 
has even parity. Therefore, the integrals are nonzero in general, provided that they are not symmetric
under $\nu\leftrightarrow\nu'$. Though the latter depends on the matrices $\{\gamma_\nu\}$, this
should be the case in general. 

Finally, only the upper boundary of the integration in Eq. (\ref{tensor2}) depends on $|m|$. Separating 
the Green's function into an even parity term $s_\bp$ and an odd parity term $a_\bp$,
these two terms have a different asymptotic behavior for large $\bp$, namely
$
a_\bp\sim p^{-1}
$,
$
s_\bp\sim p^{-2}
$.
This implies for the dependence on the upper cut-off $a$ 
\beq
\int_\bp a_\bp\partial_{p_\lambda}s_\bp\sim {\rm const.}+ a^{d-4}
.
\eeq
Thus, the tensor $C$ depends weakly on $a$ for $d<4$. This implies that the scaling of 
$\Gamma_{\lambda;\mu\mu'}$ depends weakly on $a$ and it is proportional to $\sgn(m)$.
The tensor $\Gamma_\lambda$ and the current operator in Eq. (\ref{current0}) are not defined exactly at $m=0$ due
to the jump, only the approach from $m<0$ or from $m>0$ give a well-defined current. The jump can be identified 
with the parity anomaly known from quantum electrodynamics~\cite{witten16,adler69,redlich84},
where the phonon field is replaced by the electromagnetic gauge field.


\subsection{Chiral model for 2+1 dimensional Weyl fermions}
\label{sect:weyl}

\begin{figure*}[t]
\begin{center}
\includegraphics[width=9cm]{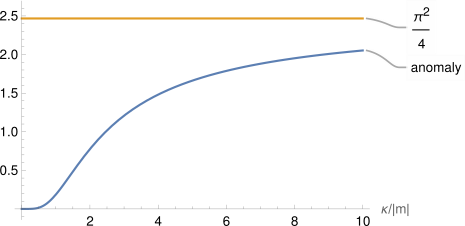}
\caption{
Phonon anomaly from the integral in Eq. (\ref{integral3})
with the asymptotic value $\pi^2/4$ for a large integral cut-off $a=\kappa/|m|$.
}
\label{fig:c}
\end{center}
\end{figure*}

As an instructive example we consider a two-dimensional Weyl Hamiltonian with odd parity 
\beq
\label{dirac01}
{\tilde H}_\bk
=k_2\sigma_2+k_1\sigma_1 
\eeq
and Pauli matrices $\{\sigma_\mu\}$.
Then with the Green's function ${\tilde G}_\bk=({\tilde H}_\bk+k_3\sigma_3+im\sigma_0)^{-1}$ we get
from Eq. (\ref{C_tensor2})
\beq
\label{C_tensor3}
C_{\lambda;\nu\nu'}
=\frac{4\pi}{3}\sgn(m)(\delta_{\nu'0}\delta_{\nu\lambda}-\delta_{\nu0}\delta_{\nu'\lambda})
\int_0^a  \frac{p^2}{(1+p^2)^3}p^2dp
\eeq
with the integral
\beq
\label{integral3}
\frac{4\pi}{3}\int_0^a  \frac{p^4}{(1+p^2)^3}dp
=\frac{4\pi}{3}\left[\frac{3}{8}\arctan a-\frac{5a^3+3a}{8(1+a^2)^2}\right]
=\frac{\pi}{2}\arctan a-\pi\frac{5a^3+3a}{6(1+a^2)^2}
.
\eeq
This yields in the limit $a\sim\infty$
\beq
C_{\lambda;\nu\nu'}\sim 
\frac{\pi^2}{4}\sgn(m)(\delta_{\nu'0}\delta_{\nu\lambda}-\delta_{\nu0}\delta_{\nu'\lambda})
,
\eeq
with the phonon parity anomaly at $m=0$. 
We note that this jump for $m\to-m$ is quite robust, it even exists for a finite cut-off $a$, as illustrated in Fig. \ref{fig:c}.
This is crucial, since for the lattice model we have a finite momentum cut-off. 
Then we get from Eq. (\ref{Gamma_tensor})
\beq
\label{Gamma4}
\Gamma_{\lambda;\mu\mu'}
=\frac{\pi^2}{4}\sgn(m)\tau_{\mu'\lambda\mu0}
=-i\pi^2\sgn(m)\epsilon_{\mu\lambda\mu'}
,
\eeq
since for Pauli matrices we have $\tau_{\mu'\lambda\mu0}=-4i\epsilon_{\mu\lambda\mu'}$.
With this result the transformation matrix in Eq. (\ref{coordinate_tr})
yields $\vK=-i\pi^2\sgn(m)\bk$ in this example. 
This implies for the eigenvectors in Eq. (\ref{eigenvectors})
\beq
\label{eigenvectors1}
{\vec\psi}_\pm=\frac{1}{\sqrt{2}\sqrt{k_2^2+k_3^2}k}\pmatrix{
k_2^2+k_3^2 \cr
-k_1k_2\mp ik_3k \cr
-k_1k_3\pm ik_2k\cr 
}
,
\eeq
while the current expectation in Eq. (\ref{current1}) becomes
\beq
({\vec\psi}_\pm\cdot\Gamma_\lambda{\vec\psi}_\pm)
=\mp\frac{\pi^2}{k}(k_1\epsilon_{2\lambda 3}-k_2\epsilon_{1\lambda3}+k_3\epsilon_{1\lambda2})
=\pm\pi^2k_\lambda/k
.
\eeq
Thus, the expectation of the current does not depend on $\sgn(m)$ in this case. 
The fermion current, on the other hand, reads $({\vec\phi}_\pm\cdot\sigma_\lambda{\vec\phi}_\pm)=\pm k_\lambda/k$.
It should be noted that this expectation values refer to the special eigenvectors of Eq. (\ref{eigenvectors1})
and the eigenvectors of the Weyl Hamiltonian $(1,\pm (k_1+ik_2)/k)^T$, respectively. However, their
corresponding eigenvalues are degenerate with respect to a rotation.
Therefore, the current expectation value depends strongly on the boundary conditions and is uniquely
determined only for the geometry of the specific system.

Since $\Gamma_{\lambda;\mu\mu'}$ is proportional to $\sgn(m)$, the current operator in Eq. (\ref{current0})
and the Berry curvature of Eq. (\ref{Berry_curvature1}) are affected by the anomaly. There are exceptions though, 
as indicated 
by the special form of $\beta_{\lambda\lambda'}$ in Eq. (\ref{current1}). This originates in the contribution of the terms 
linear in $K_2$ and $K_3$ of the eigenvectors. Other current expectation values may depend on the anomaly,
in particular, if $\psi_j$ does not depend on $\Gamma_{\lambda;\mu\mu'}$. Therefore, the general form
in Eq. (\ref{current01}) indicates that the phonon current expectation is sensitive to the anomaly.

For the physical interpretation of the anomaly it is crucial that $im$ corresponds either to absorption or to 
emission of a material, depending on its sign. This property is accessible in many materials and metamaterials,
especially for two-dimensional systems.
A brief review of recent experiments for the tuning of the absorption and emission properties can be
found in Ref.~\cite{ren21}.

\subsection{Non-chiral model with eight nodes}
\label{sect:non-chiral}

For the special case $A=\sum_{\nu=1}^3\gamma_\nu\sin k_\nu$ and $S=1$ in the fermion
Green's function $G=(A+mS)^{-1}$, we have 8 nodes for $k_\nu=0,\pi$
with the same number of positive and negative chiralities.
This gives, for instance, with $c_\nu=\cos k_\nu$ and $s_\nu=\sin k_\nu$
\beq
\label{C_tensor}
C_{2;12}=m\int_\bk\frac{c_2(s_1^2+s_3^2+m^2)}{(s_1^2+s_2^2+s_3^2+m^2)^2}
.
\eeq
Analogous expressions we get for the other values of $\lambda$ and $\nu,\nu'$.
This integral vanishes for a toroidal Brillouin zone due to the factor $c_2$. Thus, we 
have $\Gamma_{\lambda;\mu\mu'}=0$ for this case of non-chiral fermions. To get this result it is crucial that
the Brillouin zone is closed due to periodic boundary conditions. When we remove parts of the Brillouin zone
by, e.g., removing one node and its vicinity, we create edges.
The contribution of a single node and its small vicinity is easy to calculate for the integral and gives the same value 
as a single node for a Weyl fermion in Sect. \ref{sect:weyl} but with the opposite sign.
In this case we get $C_{2;12}\ne0$. This is also true when we consider a finite system in real space with edges.
In that case we have to replace the integral in Eq. (\ref{C_tensor}) by a sum, which makes the calculations more 
cumbersome though.

\subsection{Thermal fluctuations}

\begin{figure*}[t]
\begin{center}
\includegraphics[width=8cm]{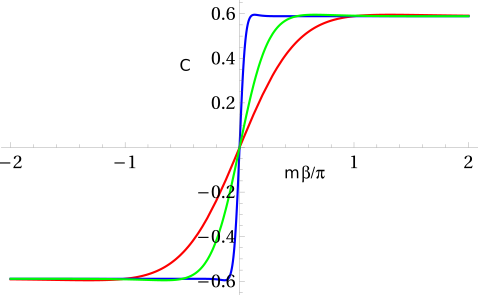}
\caption{Effect of thermal fluctuations on the integral ${\rm sgn}(m)C(m\beta)$ is a broadening of the step function
at the anomalous point $m=0$: For different temperatures the integral is visualized as a function of the
broadening $m$.  The colors indicate $\beta\equiv1/k_B T=1$ (red), $\beta=2$ (green) and $\beta=10$ (blue), where
$k_BT$ is measured in the energy units of $m$.
}
\label{fig:d}
\end{center}
\end{figure*}

So far, we have considered phonons without thermal fluctuations.
As mentioned in the context of the functional integral in Sect. \ref{sect:functional_int}, an analytic continuation 
of the time integration to the interval $[0,i\beta]$ can be used to describe thermal fluctuations. We still can
apply a Fourier transformation for the time in this case, but since the integration interval is finite, the 
corresponding frequencies are discrete; i.e., $k_3\to i\nu_n$. For the $2+1$ dimensional Weyl fermions 
in Sect. \ref{sect:weyl} these are the Matsubara frequencies $\nu_n=(2n+1)\pi/\beta$, such that the corresponding
Green's function becomes
\beq
{\tilde G}_{k_1,k_2,\nu_n}=({\tilde H}_{k_1,k_2}+i\nu_n\sigma_3+im\sigma_0)^{-1}
.
\eeq
After replacing $H_\bp$ by ${\tilde H}_{p_1,p_2}+i\nu_n\sigma_3$,  this gives for the integral in Eq. (\ref{C_tensor3})
\beq
\int_0^a  \frac{p^2}{(1+p^2)^3}p^2dp \to 
C(m\beta)= \frac{2\pi}{m\beta}
\sum_{n\ge 0}\int_0^a  \frac{p^2+\nu_n^2/m^2}{(1+p^2+\nu_n^2/m^2)^3}pdp
,
\eeq
where the integration and the summation can be calculated explicitly in terms of hyperbolic functions.
The result in Fig. \ref{fig:d} demonstrates that at low temperature the current saturates at a finite value: 
$C(m\beta)$ approaches a plateau for $m\beta\sim\pm\infty$, while for $m\beta\sim0$ the integral vanishes 
linearly. This behavior results in a broadening of the step function near the anomaly at $m=0$, reflecting a 
continuous current in Eq. (\ref{Gamma4}):
\beq
\Gamma_{\lambda;\mu\mu'}
=-\frac{16\pi i}{3}\sgn(m)C(m\beta)\epsilon_{\mu\lambda\mu'}
,
\eeq
where the slope of $\sgn(m)C(m\beta)$ as a function of $m$ increases with $1/k_BT$ at $m=0$.  
The plateau values are not affected by the thermal fluctuations, though.
For conventional phonon transport it is known that the longitudinal heat current scales at higher temperature 
as $\sim T^2$, while the transverse current does not exist~\cite{mingo05,jiang09}. This scaling persist as long 
as phonon-phonon scattering is neglected. In our approach such a behavior is caused by the second order 
derivatives of the phonon fluctuations in Eq. (\ref{gaussian0}), which are not included in Eq. (\ref{cs_action}).

\section{Conclusion}
\label{sect:conclusion}

This work has demonstrated how the interaction between chiral fermions and local, dispersionless phonons
induces nontrivial topological features in the phononic sector. Central to our analysis is the coordinate transformation 
in Eq. (\ref{coordinate_tr}), parametrized by the set of coefficients $\Gamma_{\lambda;12}$, $\Gamma_{\lambda;13}$, 
and $\Gamma_{\lambda;23}$, where
$\lambda=1,2,3$ labels two spatial and one temporal directions. This transformation leads to a clear separation 
of the fermionic and phononic dynamics, and facilitates the transfer of geometric properties, in particular the Berry 
curvature, from the electronic to the phononic degrees of freedom.
As a direct outcome of this mapping, the effective phonon Green's function acquires a Chern-Simons term, provided
the fermion Green's function breaks parity in a way that renders it neither even nor odd. 
$\{\Gamma_{\lambda;\mu\mu'}\}$ encodes topological
characteristics originally present in the fermionic system. On compact manifolds such as the toroidal Brillouin zone, 
however, the total Chern-Simons contribution vanishes due to pairs of singularities with opposite chirality.

The presence of a non-vanishing Chern-Simons term is the manifestation of a phonon parity anomaly, which yields
an observable discontinuity 
in the phonon current. This anomaly is rooted in the singular structure of the fermionic Green’s function along the real 
axis, illustrating a deep connection between fermionic singularities and phononic anomalies. 
The phonon spectrum emerging from this interaction consists of three bands: a dispersionless flat band and two bands
with linear dispersion. All three converge at a nodal point at $\vK = 0$, where the spectrum is degenerate. While the flat 
band exhibits vanishing Berry curvature, the upper and lower bands host oppositely signed curvature contributions with
hedgehog-like vector profiles, indicating the formation of effective monopole configurations in momentum space.
The effect of the phonon parity anomaly survives thermal fluctuations in the form of Hall plateaux with
a continuous transition. 

In this paper we have focused on the transverse phonon currents. Our approach in Sect. \ref{sect:functional_int}
could also be used to determine the electronic currents after integrating out the phonon degrees of freedom.
This leads to an effective theory for the electrons and their transport properties.
The longitudinal and transverse optical conductivities were calculated previously for massive Dirac fermions with
strong spin–orbit scattering and electron-phonon coupling~\cite{li13}. For monolayer phosphorene, a material
with broken inversion symmetry, the effect of theelectron-phonon interaction on the electron conductivity
was studied recently, where a vanishing transverse conductivity was found~\cite{yar24}. 

Our results indicate that phonon currents can serve as sensitive probes of the chirality and topological features 
of the underlying fermionic sector. Following the same line of reasoning, we anticipate a similar inheritance of 
geometric and topological properties of spin systems for spin-coupled phonons. 
The phonon chirality itself can be measured either through the transverse heat transport~\cite{li20a}
or more directly by the absorption of polarized infrared photons \cite{zhu18}.

Future work could consider specific fermionic or spin Hamiltonians and incorporate a specific phonon 
dispersion through spatially dependent couplings $\omega_{\br\br'}$. Additionally, our results suggest that geometric 
and boundary effects, such as edge states and finite-size responses, could play a significant role in shaping observable
phonon transport properties. Such extensions may provide valuable insights for the theoretical model and 
the experimental realization of topologically engineered phononic systems.

\appendix

\section{Linear differential term}
\label{app:CS}

The Gaussian fluctuation term in Eq. (\ref{gaussian0}) reads
\[
\int_{\bx,\bx'}Tr_{N}(q_\bx G_{\bx-\bx'}q_{\bx'}G_{\bx'-\bx})
=\int_{{\bar\bx},\bx}Tr_{N}(q_\bx G_{{\bar\bx}}q_{\bx-{\bar\bx}}G_{-{\bar\bx}})
\]
with ${\bar\bx}=\bx-\bx'$. 
The expansion of $q_{\bx-\bx'}$ around $\bx$ gives in the leading order
\[
q_{\bx-{\bar\bx}}\sim q_\bx-\sum_{\lambda=1}^3{\bar x}_\lambda\frac{\partial q_\bx}{\partial x_\lambda}
\]
such that we can write 
\beq
\label{2PGF01}
\int_{{\bar\bx},\bx}Tr_{N}(q_\bx G_{{\bar\bx}}q_{\bx-{\bar\bx}}G_{-{\bar\bx}})
\]
\[
\sim 
\int_{{\bar\bx},\bx}Tr_{N}(q_\bx G_{{\bar\bx}}q_{\bx}G_{-{\bar\bx}})
-\int_{{\bar\bx},\bx}\sum_{\lambda=1}^3{\bar x}_\lambda
Tr_{N}(q_\bx G_{{\bar\bx}}\frac{\partial q_\bx}{\partial x_\lambda}G_{-{\bar\bx}}
) + {\rm higher}\ {\rm order}\ {\rm gradient}\ {\rm terms}
\ .
\eeq
With the partial integration 
\beq
\int_\bx
Tr_{N}\left(q_\bx G_{{\bar\bx}}\frac{\partial q_\bx}{\partial x_\lambda}G_{-{\bar\bx}}\right)
=-\int_\bx
Tr_{N}\left(\frac{\partial q_\bx}{\partial x_\lambda}G_{{\bar\bx}}q_\bx G_{-{\bar\bx}}\right)
,
\eeq
where we neglected the surface term, we get from Eq. (\ref{2PGF01})
\beq
\label{2PGF02}
\int_{{\bar\bx},\bx}Tr_{N}(q_\bx G_{{\bar\bx}}q_{\bx-{\bar\bx}}G_{-{\bar\bx}})
\]
\[
\sim 
\int_{{\bar\bx},\bx}Tr_{N}(q_\bx G_{{\bar\bx}}q_{\bx}G_{-{\bar\bx}})
-\frac{1}{2}\int_{{\bar\bx},\bx}\sum_{\lambda=1}^3{\bar x}_\lambda
\left[
Tr_{N}(q_\bx G_{{\bar\bx}}\frac{\partial q_\bx}{\partial x_\lambda}G_{-{\bar\bx}})
-Tr_{N}(\frac{\partial q_\bx}{\partial x_\lambda}G_{{\bar\bx}}q_\bx G_{-{\bar\bx}})
\right]
.
\eeq
Finally, an expansion 
of $q_\bx$, $G_\bx$ in terms of $\{\gamma_\mu\}$ matrices yields for the linear differential term
\beq
\label{antisymmetric1}
-\frac{1}{2}\int_{{\bar\bx},\bx}\sum_{\lambda=1}^3{\bar x}_\lambda
\sum_{\mu,\mu'=1}^3
q_{\bx\mu}\frac{\partial q_{\bx\mu'}}{\partial x_\lambda}
\sum_{\nu,\nu'=0}^3G_{{\bar\bx}\nu}G_{-{\bar\bx}\nu'}
\left[Tr_{N}(\gamma_\mu\gamma_\nu\gamma_{\mu'}\gamma_{\nu'})
-Tr_{N}(\gamma_{\mu'}\gamma_\nu\gamma_{\mu}\gamma_{\nu'})\right]
.
\eeq


\end{document}